\documentclass[12pt]{article}
\usepackage{cite,epsfig}

\def\BA{\begin{eqnarray}}
\def\BE{\begin{equation}}
\def\EA{\end{eqnarray}}
\def\EE{\end{equation}}

\def\Address#1#2{$^{\rm#1}${\it\footnotesize#2}\\}
\def\Ref#1{(\ref{#1})}
\def\cf{{\it cf. }}
\def\eps{\varepsilon}
\def\Dtau{\Delta\tau}
\def\absP{\langle\vert P\vert\rangle}
\def\gtsim{\lower-0.45ex\hbox{$>$}\kern-0.77em\lower0.55ex\hbox{$\sim$}}
\def\ltsim{\lower-0.45ex\hbox{$<$}\kern-0.77em\lower0.55ex\hbox{$\sim$}}

\begin{document}

\title{\bf Transverse QCD Dynamics \\ Near the Light Cone}
\author{
  E.--M.~Ilgenfritz$^{\rm a,c}$,
  Yu.P.~Ivanov$^{\rm a,b,d}$,
  H.J.~Pirner$^{\rm a,b}$\footnote{
    Supported by the European Contract No. FMRX-CT96-0008.
  }
  \\
  \\
  \Address{a}{
    Institut f\"ur Theoretische Physik der Universit\"at,
    Philosophenweg 19, D-69120 Heidelberg, Germany
  }
  \Address{b}{
    Max-Planck Institut f\"ur Kernphysik,
    Postfach 103980, D-69029 Heidelberg, Germany
  }
  \Address{c}{
    Research Center for Nuclear Physics,
    Osaka University, Osaka 567-0047, Japan
  }
  \Address{d}{
    Joint Institute for Nuclear Research,
    Dubna, 141980 Moscow Region, Russia
  }
}
\date{\today}
\maketitle

\begin{abstract}
  Starting from the QCD Hamiltonian in near-light cone coordinates,
  we study the dynamics of the gluonic zero modes. Euclidean 2+1 
  dimensional lattice simulations show that the gap at strong 
  coupling vanishes at intermediate coupling. This result opens
  the possibility to synchronize the continuum limit with the 
  approach to the light cone.
\end{abstract}

\section{Introduction}

The solution of Quantum Chromodynamics {\em on} the light cone
is still an unsolved theoretical task the present status of which
is reviewed in ref. \cite{Brodsky}. In a recent paper \cite{NPFV} 
a formulation of QCD in coordinates {\em near} the light cone 
has been proposed which has the advantage of keeping a direct 
link \cite{PRFR,LTLY,VFP} to equal time theories. The Cauchy 
problem is well defined in near light cone coordinates, since 
the initial data are given on a space like surface. This formulation
avoids the solution of constraint equations which, on the quantum
field theoretic level, may be very complicated. The problem of 
a nontrivial vacuum appears in a solvable form related to the 
transverse dynamics. This is physically very appealing, since 
in high energy reactions the incoming particles propagate near the
light cone and interact mainly exchanging particles with transverse
momenta. We would like to connect successful models for the soft 
transverse nonperturbative dynamics of high energy reactions 
(\cf refs. \cite{DGKP,N}) to the underlying QCD Hamiltonian. 
In diffractive reactions, fast Lorentz contracted hadrons 
experience the confining forces of QCD in their transverse 
extensions and interact with other fast moving hadrons via 
soft interactions. In a theory of total cross sections the 
nonperturbative infrared dynamics in the transverse plane is
essential. The objective of this paper is to investigate the
effective transverse Hamiltonian in QCD near the light cone
for small light cone momenta.
        
Near light cone QCD has a nontrivial vacuum which we claim
cannot be neglected even in the light cone limit. We will 
demonstrate the existence of massless excitations in the 
zero mode theory. These excitations do not decouple in the
light cone limit. Genuine nonperturbative techniques must be
used to investigate the behavior of this limit. In principle
the additional parameter which labels the coordinate system
can be chosen arbitrarily. We will show that the zero mode
Hamiltonian depends on an effective coupling constant containing
this parameter and evolves towards an infrared fixed point. 
Therefore, we propose a sophisticated choice of the frame 
dependence which facilitates calculations in the near light
cone frame. We use the infrared fixed point of the zero mode
Hamiltonian to follow a trajectory in the space of couplings,
where high resolution is synchronized with the light cone limit.
Note, we consider the parameter associated with the frame 
dependence as yet another coupling constant. As shown in 
Ref. \cite{Pol} the zero mode sector is also relevant
to the definition of M-theory in light cone coordinates. 

We choose the following near light cone coordinates which
smoothly interpolate between the Lorentz and light front
coordinates:
\BA
  x^t ~=~ x^{+} &=& \frac1{\sqrt2} 
    \left\{ \left(1 + \frac{\eta^2}{2} \right)
    x^{0} + \left(1 - \frac{\eta^2}{2} \right) x^{3} 
    \right\} ~,\nonumber \\
          x^{-} &=& \frac1{\sqrt2} \left( x^{0}-x^{3} \right)~.
\label{Coor}
\EA
The transverse coordinates $x^1$, $x^2$ are unchanged;
$x^t = x^{+}$ is the new time coordinate, $x^{-}$ is a 
spatial coordinate. As finite quantization volume we will 
take a torus and its extension in ``-'', as well as in ``1, 2'' 
direction is $L$. The scalar product of two 4-vectors $x$ and $y$
is given with $\vec{x}_\bot \vec{y}_\bot = x^1 y^1 + x^2 y^2$ as
\BA
  x_\mu y^\mu & = & x^- y^+ + x^+ y^- - \eta^2 x^- y^-
                - \vec{x}_\bot \vec{y}_\bot \nonumber \\
              & = & x_- y_+ + x_+ y_- + \eta^2 x_+ y_+
                - \vec{x}_\bot \vec{y}_\bot ~.
\label{scalpr}
\EA
Obviously, the light-cone is approached as the parameter $\eta$ 
goes to zero. For non-zero $\eta$, the transition to the coordinates
introduced above can be formally identified as a Lorentz-boost 
combined with a linear transformation, which avoids time dependent
boundary conditions \cite{LTLY}. The boost parameter $\beta = v_3$
is given by
\BE
  \beta = \frac{1 - \eta^2/2}{1+ \eta^2/2} ~,
\label{beta}
\EE
indicating that for $\eta^2 \rightarrow 0$ the relative velocity
$v_3 \rightarrow c\, (\equiv 1)$. Of course, this is connected to the
well-known interpretation of the 'tilted' light-cone frame in terms 
of the infinite momentum frame.

The choice of near light cone coordinates allows to quantize the
theory on a space-like finite interval of length $L$ in $x^-$ at
equal times, {\it i.e.} $\Delta x^+ = 0$. The invariant length 
squared of this interval for $\Delta x_{\bot}^2=0$ is related
to the length of the compact $x^-$ dimension.
\BA
  \Delta s^2 & = & \Delta x^- \Delta x^+ + \Delta x^+ \Delta x^-
               - \eta ^2 (\Delta x^-)^2 - \Delta x_\bot^2 \nonumber \\
             & = & -\eta^2 L^2 ~.
\EA
For simplicity we consider also the transverse dimensions periodic 
in $L$. Previous work of the St.~Petersburg and Erlangen groups 
\cite{PRFR,LTLY} assumed a fixed tilted coordinate system with 
fixed transverse ultraviolet cut off. Our main purpose in this
paper is to consider the zero mode fields on a transverse lattice 
with {\it varying} transverse lattice spacing $a$. We propose to
approach light-cone dynamics by synchronizing the continuum limit
$\Lambda=\pi/a\to\infty$ with the light-cone limit $x^+\to\frac{1}%
{\sqrt 2}(x^0+x^3)$. The vacuum fluctuations in the transverse 
directions induce a second order phase transition which allows 
to have a continuum limit of the lattice theory. Thereby we can 
eliminate the cut off in a controlled way, preserving the
nontrivial vacuum structure. 

The gauge fixing procedure in the modified light-cone gauge 
$\partial_-A_- = 0$ involves zero modes dependent on the transverse 
coordinates. These zero mode fields carry zero linear momentum $p_-$
in near light cone coordinates, but finite amount of $p_0+p_3$.
They correspond to "wee" partons in the language of the original
parton model of Feynman. In $SU(2)$ the zero mode fields $a_-(x_\perp)$
are proportional to $\tau^3$, i.e. they can be chosen color diagonal. 
The use of an axial gauge is very natural for the light-cone 
Hamiltonian even more so than in the equal-time Hamiltonian.
The asymmetry of the background zero mode naturally coincides 
with the asymmetry of the space coordinates on the light cone. 
The zero mode fields describe disorder fields. Depending on the
effective coupling the zero mode transverse system will be in the
massive or massless phase. Its second order phase transition allows
to perform the continuum limit in the zero mode Hamiltonian. Our 
main conjecture is that the continuum limit and light cone limit 
can be realized simultaneously at this critical point. The evolution
of the coupling determines the approach to the light cone with 
transverse resolution approaching zero. The resulting relation 
is reminiscent of the simple behavior one gets from considering
naive scaling relations and dynamics in the infinite momentum 
frame. We will also calculate the contribution of the two dimensional 
zero modes to the ground state energy and show that it scales with the
three dimensional volume. This estimate demonstrates very simply the
relevance of modes with lower dimensionality to the full problem in
the case of critical second order behavior. 

The naive considerations for a simultaneous light cone and continuum
limit go as follows. We characterize the 'infinite' momentum frame 
by giving the momenta of the proton and photon in usual coordinates.
The fast moving proton carries $(P,0_{\bot},P)$ and the photon
$q= (\frac{\nu M}{2 P},\sqrt{Q^2},-\frac{\nu M}{2 P})$, where
$\nu$ is the energy transfer in the laboratory. The dominant 
contribution comes from energy conserving transitions, where 
the energy of the quark in the final state equals the sum of 
initial quark and photon energies. For collinear quarks with 
momentum $p=(x_B P,0_{\bot},x_B P)$, where $x_B$ denotes the 
momentum fraction, one explicitly finds
\BE
  \sqrt{Q^2+\left(x_B P-\frac{\nu M}{2 P}\right)^2}=x_B P+\frac{\nu M}{2P},
\EE
which yields the ``scaling variable'' $x_B=\frac{Q^2}{2 M \nu}$
as long as $P\geq Q/x_B$, i.e. the ``infinite'' momentum has to be 
large enough:
\BE
  \gamma M = P \geq Q = 1/a.
\EE
At the same time, the photon has a wavelength $\propto 1/Q$ which
resolves transverse details of size $a$ in the hadronic wavefunction.
By eq. \Ref{beta}, the $\gamma$ factor of the ``infinite'' momentum
is related to $\eta$. Therefore we expect that the parameter $\eta$ 
approaches zero with the transverse cutoff $\Lambda=\pi/a\rightarrow\infty$
\BE
  \eta=\frac1{\sqrt{2}\gamma} \approx \frac{1}{\sqrt{2}}\,aM \rightarrow 0.
\EE
One of the objectives of this paper is to derive the precise relation
between $\eta$ and $a$. We will simulate the zero mode dynamics 
on a $(2+1)$ dimensional lattice to find the fixed point of the 
effective coupling. The light-like limit is governed by an infrared
fixed point (\cf ref. \cite{Pol}). Most previous discussions of the
zero mode problem have been on the classical equations of motion level, 
here we will work on the fully nonperturbative quantum level.
Our work is not directly related to M-theory, but some of the 
consequences may be relevant for other studies with light like 
compactification.

\section{Near Light Cone QCD Hamiltonian} 

In Ref. \cite{NPFV} the near light cone Hamiltonian has been derived.
A similar Hamiltonian has been obtained in equal time coordinates
\cite{LNOT,LNT}. Here we will sketch the derivation. We restrict
ourselves to the color gauge group $SU(2)$ and dynamical gluons;
only an external (fermionic) charge density $\rho_m$ is 
considered here.

Since the $A^a_+$ coordinates have no momenta conjugate to them,
the Weyl gauge $A^a_+=0$ is the starting point for a canonical
formulation. The canonical momenta of the dynamical fields 
$A_-^a, A_i^a$ are given by
\BA
  \Pi^a_- &=& \frac{\partial{\cal L}}{\partial F^a_{+-}} = F^a_{+-}, 
  \nonumber \\
  \Pi^a_i &=& \frac{\partial{\cal L}}{\partial F^a_{+i}} = F^a_{-i}
             +\eta^2 F^a_{+i}.
\EA
>From this, we get the Weyl gauge Hamiltonian density
\BE
  \label{Hweyl}
  {\cal H}_W = \frac{1}{2} \Pi^a_- \Pi^a_-
    + \frac{1}{2}F^a_{12}F^a_{12}
    + \frac{1}{2\eta^2}\sum_{i=1,2}\left(\Pi^a_i-F^a_{-i}\right)^2 ~.
\EE
The Hamiltonian has to be supplemented by the original Euler--Lagrange 
equation for $A_+$ as constraints on the physical states (Gauss' Law
constraints)
\BA
  \label{Gauss}
  G^a({x}_{\bot},x^-) |\Phi\rangle
  &=& \left( D_-^{ab} \Pi^b_{-}
           + D_{\bot}^{ab}\Pi^b_{\bot}
           + g \rho_m^a
      \right) | \Phi \rangle \nonumber \\
  &=& \left(D_-^{ab}\Pi^b_{-} + G_{\bot}^{a}\right) |\Phi\rangle = 0~.
\EA

In order to obtain a Hamiltonian formulated in terms of unconstrained
variables, one has to resolve the Gauss' Law constraint. Via unitary
gauge fixing transformations [9,10] a solution of Gauss' Law with 
respect to components of the chromo--electric field $\Pi_{-}$ can be
accomplished. This gives a Hamiltonian independent of the conjugate
gauge fields $A_{-}$, {\it i.e.} the latter become cyclic variables. 
Classically this would correspond to the light front gauge $A_{-}=0$. 
However, this choice is not legitimate if we want to consider the 
theory in a finite box. Instead, the (classical) Coulomb light front
gauge $\partial_{-}A_{-}=0$ is compatible with gauge invariance and
periodic boundary conditions. The reason is that $A_{-}$ carries 
information on the (gauge invariant) eigenvalues of the spatial 
Polyakov line matrix 
\BE
  \label{Polyakov}
  \hat{\cal P}(x_\bot)=P\exp\left[ig\int dx^-A_-(x_\bot,x^-)\right] ~,
\EE
which can be written in terms of a diagonal matrix with the zero mode
field $a^3_-(x_\bot) \frac{\tau_3}{2} = a_-(x_\bot)$
\BE
  \hat{\cal P}(x_\bot) = 
  V \exp\left[ig L a_-(x_\bot)\right] V^{\dagger} ~.
\EE
Obviously we have to keep these `zero modes' $a_-(x_\bot)$ as 
dynamical variables, while the other components of $A_{-}$ are
eliminated. The zero mode degrees of freedom are independent of
$x^{-}$ and, therefore, correspond to quantities with zero 
longitudinal momentum $p_{-}$.

In order to eliminate the momentum $\Pi_-$, conjugate to $A_-$, by 
means of Gauss' Law, one needs to `invert' the covariant derivative
$D_-$ which simplifies to $d_{-}=\partial_{-} - ig[a_{-},..]$. On 
the space of physical states one can simply make the replacement%
\footnote{The inversion of $d_-$ can be explicitly performed in 
terms of its eigenfunctions, \cf \Ref{Gauss}.}
\BE
  \Pi_-(x_\bot,x_-) \rightarrow
    p_-(x_\bot)-\left(d_-^{-1}\right) G_\bot(x_\bot,y^-) ~.
\EE
The operator $p_-({x}_{\bot})$ is also diagonal and $p^3_-({x}_{\bot})$
is the momentum conjugate to the zero mode $a^3_{- }(x_{\bot})$.  It has 
eigenvalue zero with respect to $d_{-}$, {\it i.e.} $d_- p_- = 0$, and
is therefore not constrained.

The appearance of the zero modes implies a residual Gauss' Law, which 
arises from the $x_{-}$ integration over eq. \Ref{Gauss} for the $a=3$
component using periodic boundary conditions in the $x_{-}$ direction. 
This constraint on two dimensional fields can be handled in full analogy
to QED, since it concerns the diagonal part of color space. A further 
Coulomb gauge fixing in the $SU(2)$ 3--direction eliminates the color
neutral, $x^-$--independent, two--dimensional longitudinal gauge field
in favor of a neutral chromo--electric field
\BE
  e_\bot(x_\bot) = g \nabla_\bot \int dy^- dy_\bot
    d(x_\bot - y_\bot)
    \left\{ f^{3ab} A^a_\bot(y_{\bot}, y^-) \Pi^b_{\bot}(y_{\bot}, y^-)  
    + \rho^3_m(y_{\bot}, y^-) \right\} \frac{\tau^3}{2} ~.
\EE
Here we use the periodic Greens function of the two dimensional 
Laplace operator
\BE
\label{Green2}
  d(z_\bot) = - \frac1{L^2} \sum_{\vec{n} \neq \vec0 }
  \frac1{p_n^2} e^{i p_n z_\bot}    ~,
  ~~~~~ p_n = \frac{2\pi}{L}\vec{n} ~,
\EE
where $\vec{n} = (n_1,n_2)$ and $n_1,n_2$ are integers.  

As a remnant of the local Gauss' Law constraints, a global condition  
\BE
  \label{Gaussgl}
  Q^3 | \Phi' \rangle = 
    \int dy^- dy_\bot
    \left\{ f^{3ab} A^a_\bot(y_\bot,y^-) \Pi^b_\bot(y_\bot, y^-)  
      + \rho^3_m(y_{\bot}, y^-) \right \} 
   | \Phi' \rangle = 0 
\EE
emerges. The physical meaning of this equation is that the neutral 
component of the total color charge, including external matter as 
well as gluonic contributions, must vanish in the sector of physical
states. It can be shown (see \cite{NPFV}) that there must be 
$\tilde{Q}_{12}(x_{\bot})=0$ everywhere in transverse space in 
order to avoid an infinite Coulomb energy. The two conditions 
together suggest that physical states have to be color singlets.  

The final Hamiltonian density in the physical sector explicitly reads
(in terms of unconstrained $A_{\bot}$ and $\Pi_{\bot}$ obtained after 
a shift is made by subtracting averages) \cite {NPFV}
\BA
  \label{Hamil}
  {\cal H} &=& \mbox{tr}
     \left[\partial_1 A_2 - \partial_2 A_1 - ig [A_1, A_2] \right]^2
   + \frac1{\eta^2} \mbox{tr} \left[\Pi_\bot - \left(\partial_-A_\bot
   - ig [a_-,A_\bot] \right) \right]^2 \nonumber \\
  &+& \frac1{\eta^2} \mbox{tr} 
     \left[ \frac1L e_\bot - \nabla_\bot a_- \right]^2 
   + \frac1{2 L^2} p_-^{3\,\dagger}(x_\bot) p^3_-(x_\bot) \nonumber \\
  &+& \frac1{L^2} \int_0^L dz^- \int_0^L dy^-
     \sum_{p,q,n}\,^{'}
       \frac{G_{\bot qp}(x_\bot,z^-)
             G_{\bot pq}(x_\bot,y^-)}{\left[ \frac{2\pi n}{L}
       + g(a_{-q}(x_\bot) - a_{-p}(x_\bot)) \right]^2}
         e^{i 2\pi n(z^- - y^-)/L} ~,
\EA
where $p$ and $q$ are matrix labels for rows and columns, $a_{-q} = %
(a_-)_{qq}$ and the prime indicates that the summation is restricted
to $n \neq 0$ if $p = q$. The operator $G_\perp\left(x_\bot,x^-\right)$
is defined as
\BE
G_{\perp}=
  \label{Gop}
  \nabla_\perp \Pi_\perp
    + gf^{abc} \frac{\tau^a}{2} A^b_\perp \left( \Pi^c_\perp
    - \frac1L e^c_\bot\right) 
    + g\rho_{m} ~.
\EE

The last two terms of the Weyl gauge Hamiltonian come from 
the original term $\Pi_-^2$ with squared electric field 
strengths in $x^-$ direction. After elimination of $\Pi_-$
the zero mode part and the light cone Coulomb energy in the
axial gauge remain. In the Coulomb term one sees the role of
the zero mode fields as infrared regulators of the spatial 
momenta $p_-=2 \pi n/ L$ which are quantized due to the compact
interval $L$. Since the two dimensional theory has been Coulomb
gauge fixed, the electric field $\vec{e}_{\bot}$ replaces the 
canonical momentum of the longitudinal, neutral gauge field 
which has been eliminated.

We note that the terms containing $\vec{e}_{\bot}$ and the 
momentum $\vec{\Pi}_{\bot}$, have the pre-factor $1/\eta^2$.
Physically this pre-factor signals the increase of transverse
electric energies with the boost factor $\gamma = (\sqrt2\eta)^{-1}$.
The boost also couples transverse electric fields with transverse 
magnetic fields. In the light-cone limit the pre-factor diverges 
and the adjacent brackets become constraint equations. This 
reflects the corresponding reduction of the number of degrees
of freedom if one goes exactly on the light cone. We do not 
follow this procedure, but keep a finite $\eta$ as a kind of
Lagrange parameter.

A characteristic feature of an exact light-cone formulation is 
the  triviality of the ground state. This may simplify explicit
calculations, {\it e.g.} of the hadron spectrum. However, the light 
cone vacuum is definitely not trivial in the zero mode sector.
In fact, already in ref \cite{NPFV} we have shown strong and 
weak coupling solutions of the zero mode Hamiltonian. Massless
modes influence the dynamics in the light cone limit, whereas
massive modes decouple when $\eta \to 0$. In contrast to earlier 
work, we solve the zero mode Hamiltonian in this paper numerically
showing the transition from the massive phase to the massless phase.

The zero mode degrees of freedom $a_-$ couple to the transverse 
three dimensional gluon fields $A_i$ via the second magnetic term
in $\cal H$, the Coulomb term and directly via the electric field
$\vec{e}_{\bot}$. We remark that the transverse electric fields 
$\Pi_\bot$ and $\vec{e}_\bot$ are dual to the magnetic fields
$\partial_-A_\bot$ and $\nabla_\bot a_-$. This duality is typical
for the light cone and is absent in the equal time case. Since 
duality plays an important role in supersymmetric QCD its role
in light cone theories should deserves to be investigated in 
greater detail. 

In the following  we neglect the couplings between the three and
two dimensional fields and consider the pure zero mode Hamiltonian
\BE
  \label{Hamil3}
  h = \int d^2\!x \; \left[ 
    \frac1{2 L} p_-^{3\dagger}(\vec{x}_\bot) p_-^3(\vec{x}_\bot)  
  + \frac{L}{2\eta^2} (\nabla_\bot a_-^3)^2 \right] ~.
\EE
The global constraint, eq. \Ref{Gaussgl}, does not contain $a_-$
and $p_-$ and, consequently, is irrelevant for the time being. 
Even at this level of severe approximations the zero mode
Hamiltonian differs from the corresponding one in QED. The
reason is the hermiticity defect of the canonical momentum 
$p^-$. In the Schr\"odinger representation eq. \Ref{Hamil3}
\BE
  \label{Hamil4}
  h = \int d^2x \; \left[ 
    - \frac{1}{2L}\frac{1}{J(a_-^3(\vec{x}_\bot))}
      \frac\partial{\partial a_-^3(\vec{x}_\bot)} 
      J\left(a_-^3(\vec{x}_\bot)\right)
      \frac\partial{\partial a_-^3(\vec{x}_{\bot})}
    + \frac{L^2}{2\eta^2} (\nabla_\bot a_-^3)^2 \right] ~,
\EE
contains the Jacobian $J(a_-)$ which equals the Haar measure of $SU(2)$
\BE
  \label{Jacob}
  J\left(a_-^3(\vec{x}_{\bot})\right) = 
   \sin^2 \left( \frac{gL}{2}a_-^3(\vec{x}_{\bot})\right) ~.   
\EE
It stems from the gauge fixing procedure, effectively introducing 
curvilinear coordinates. It also appears in the functional
integration volume element for calculating matrix elements.
It is convenient to introduce dimensionless variables 
\BE
  \varphi(\vec{x}_\bot) = \frac{gL a_-^3(\vec{x}_\bot)}{2} ~,
\EE
which vary in a compact domain $0 \leq \varphi \leq \pi$. We 
regularize the above Hamiltonian $h$ by introducing a lattice 
spacing $a$ between transversal lattice points $\vec{b}$. 
Next we appeal to the physics of the infinite momentum frame and
factorize the reduced true energy from the Lorentz boost factor
$\gamma=\sqrt2/\eta$ and the cut off by defining $h_{\rm red}$
\BE
  \label{hdef}
  h = \frac1{2\eta a} h_{\rm red}
\EE
In the continuum limit of the transverse lattice theory we let
$a$ go to zero. For small lattice spacing we obtain the reduced
Hamiltonian
\BE
  h_{\rm red} = \sum_{\vec{b}}\left\{ -g^2_{\rm eff}
    \frac1J \frac\partial {\partial \varphi(\vec b)}
          J \frac\partial {\partial \varphi(\vec b)}
   + \frac1{g^2_{\rm eff}} \sum_{i=1,2}
    \left(\varphi(\vec{b})-\varphi(\vec{b}+\vec{e}_i)\right)^2
  \right\} ~.
\EE
with the effective coupling constant
\vskip-5mm
\BE
  \label{g2eff}
  g^2_{\rm eff} = \frac{g^2L\eta}{4a}
\EE
The first part of the Hamiltonian contains the kinetic (electric) 
energy of the $SU(2)$ rotators on a half circle at each lattice 
point and the second part gives the potential (magnetic) energy
of these rotators due to the differences of angles at nearest 
neighbor sites. For further discussion we define these electric
and magnetic parts as:
\BA
  h_{\rm red} & = & h_{\rm e} + h_{\rm m} \\
  h_{\rm e}   & = & \sum_{\vec{b}} h^{\rm e}_{\vec{b}} \nonumber \\ 
  h_{\rm m}   & = & \sum_{\vec{b}}\sum_{i=1,2} 
                  h^{\rm m}_{\vec{b},\vec{b}+\vec{e}_i} \nonumber
\EA
In the strong coupling domain where we have analytical solutions
of the zero mode Hamiltonian, the numerical Hamiltonian lattice
theory agrees with the analytical solutions for the mass gap. 
The real question concerns the continuum limit of the zero mode
system which occurs outside of the strong coupling region. For 
this purpose we have to find the region of vanishing mass gap
for the lattice Hamiltonian.

\section{Lattice Calculation of the Zero Mode Hamiltonian}

We solve the equivalent lattice theory in a Euclidean formulation.
The zero mode Hamiltonian represents a $2+1$ dimensional theory in
two spatial and one time direction. The lattice has $N_x a = N_y a$
extensions in transverse space and $N_T \Dtau = T$ extension in near
light cone time. To set up the density matrix one has to write down
the Trotter formula for the given Hamiltonian. Using $h_{\rm red} =%
h_{\rm e}+h_{\rm m}$ we have
\BE
  \exp(-T h_{\rm red}) = \lim_{N_T\to\infty} 
   \left[ \exp\left(-\Dtau h_{\rm m}/2 \right)
          \exp\left(-\Dtau h_{\rm e}   \right)
          \exp\left(-\Dtau h_{\rm m}/2 \right)
   \right]^{N_T},
\EE
where each time evolution step $\Dtau$ can be separately done for
the electric and magnetic part of the Hamiltonian. For definiteness,
we choose $\Dtau=a/2$ in all the following. In the Appendix we show
that this choice of $\Dtau$ optimizes the updating procedure since
it generates approximately equal widths for the weight functions
resulting from the kinetic and potential energies. The different time
slices will labeled with the index $l$. The electric Hamiltonian can
be evaluated by inserting products of complete set single site
eigenfunctions $C_{n_l}$ (\cf \cite{NPFV}). Practically a maximal 
number $N_{\rm max} = 100$ of eigenfunctions is fully sufficient 
to reach convergence in the interval of couplings we need.
\BE
  \label{sum_over_states}
  \langle \varphi_{l+1} | h^{\rm e} | \varphi_l \rangle =
    \!\!\sum_{n_{l+1},n_{l}}
         \langle \varphi_{l+1} | n_{l+1} \rangle 
         \langle n_{l+1} | h^{\rm e} | n_l \rangle
         \langle n_l | \varphi_l \rangle \approx
    \!\!\sum_{n_{l}=0}^{N_{\rm max}} 
         C_{n_l}(\varphi_{l+1}) g_{\rm eff}^2~n_l(n_l+2) 
         C_{n_l}(\varphi_l)
\EE
with the (single site) eigenfunctions and eigenvalues given as:
\BA
  \label{Basis}
  C_{n_l}(\varphi_l) &=&
    \sqrt{\frac2\pi} \left\{
    \frac{\sin\left((n_l+1)\varphi_l\right)}{\sin\varphi_l}
    \right\} ~, \\
  h^{\rm e}~C_{n_l}(\varphi_l) &=&
    g^2_{\rm eff}~n_l (n_l+2)~C_{n_l}(\varphi_l) ~.
\EA
The eigenfunctions form an orthonormal set with respect to a scalar
product which contains the Jacobian in the measure. The Jacobian 
$J\left(\varphi(\vec b)\right)$ has been defined above \Ref{Jacob}. 
The magnetic part is diagonal in $\{\varphi(\vec{b})\}$ 
{\it i.e.} local in time:
\BE
  \langle \{\varphi_l(\vec{b})\}|h_{\rm m}|\{\varphi_l(\vec{b})\}\rangle
   = h_{\rm m}(\{\varphi_l(\vec{b})\})
   = \sum_{\vec{b}}\sum_{i=1,2}
     h^{\rm m}_{\vec{b},\vec{b}+\vec{e}_i}
     \left(\varphi_l(\vec{b})-\varphi_l(\vec{b}+\!\vec{e}_i\,)\right) ~.
\EE
The full partition function is given by an integral over all time slices
\BA
 \!\!\!Z &=&
    \mbox{tr}\,\exp\!\left[-T h_{\rm red}\right]
    = \int \!\prod_{\vec{b}} 
         \left(J(\varphi(\vec{b})) d\varphi(\vec{b})\right) \,
      \langle \{\varphi(\vec{b})\} | \exp\!\left[-T h_{\rm red}\right]
            | \{\varphi(\vec{b})\}
      \rangle \\
   &=&
     \int \!\prod_{\vec{b},l} 
         \left(J(\varphi(\vec{b})) d\varphi(\vec{b}) \,
       \sum_{n_l=0}^{N_{\rm max}}
       C_{n_l}\!(\varphi_{l+1}(\vec{b}))
       \exp\left[ -g^2_{\rm eff} n_l (n_l\!+\!2) \Dtau \right]  
       C_{n_l}\!(\varphi_{l}(\vec{b})) \right)
       \nonumber \\
     &\times&
       \prod_{l} \exp\!\left[ -h_{\rm m} (\{\varphi_l(\vec{b})\}) \Dtau \right] ~.
       \nonumber
\EA
Because of the Jacobian the dominant contributions to the partition
function come from $\varphi_l$ values around $\frac{\pi}{2}$. The
Hamiltonian is invariant under reflections $\varphi_l-\frac{\pi}{2}
\rightarrow \frac{\pi}{2}-\varphi_l$. In the strong coupling limit,
where $g^2_{\rm eff}$ is large, the half rotators act almost independently 
on each lattice site producing a large mass gap. The system is disordered.
For decreasing coupling constant $g^2_{\rm eff}$ the movement of the 
individual rotators becomes locked from one site to the next. Long 
range correlations develop. The order parameter (or ``magnetization'')
of the system is the expectation value of the trace of the Polyakov
line \Ref{Polyakov}, which on the lattice has the form
\BE
  P = \overline{\frac{1}{2} tr \hat{\cal P}}
    = \frac1{N^3}\sum\limits_{\vec b,\tau} \cos\varphi(\vec b,\tau) ~.
  \label{OrderPar}
\EE

The order parameter is odd under the above symmetry operation.
Operator expectation values are evaluated with the density matrix 
defined by $h_{\rm red}$ and $T$:
\BE
  \langle O \rangle = 
  \frac1Z\,\mbox{tr}\left\langle O\exp[-T h_{\rm red}]\right\rangle~.
\EE

At each particular $\beta_g$
\BE
  \beta_g \equiv 1/g_{\rm eff}^2
\EE
we simulate lattice sizes with equal number of sites in all directions
$N = N_x = N_y = N_T$ and $N = 4,6,8,12,16$. A Metropolis algorithm is
used for updating with the steps size and the number of hits adapted 
to $\beta_g$. In order to tabulate the Boltzmann weights related to 
the timelike links we also discretized the continuous angle variables
to a system of $N_\varphi=60$ orientations: $\varphi_j=j\pi/N_\varphi$,
$j=[0,N_\varphi]$.

In this explorative investigation we generated between $5000$ and 
$50000$ uncorrelated configurations depending on the $\beta_g$ and 
lattice size. All our calculations were done on a cluster of 
AlphaStations with the single site Metropolis updating algorithm
sketched above, so the accuracy can be improved using a more 
powerful algorithms and/or more computing resources.

We calculate the $\beta_g$ dependence of the following quantities:
average electric energy $\eps_{\rm e}$ and magnetic energy $\eps_{\rm m}$
per site, the average of the absolute value of the order parameter
$\absP$, the susceptibility $\chi$ and the normalized fourth cumulant
$g_r$ (which gives the deviation of the moments of the Polyakov 
expectation value from a pure Gaussian behavior):
\vskip-7mm
\BA
  \eps_{\rm e,m} &=& \frac1{N^2} \langle h_{\rm e,m} \rangle \\
  \chi           &=& N^3 \left(\langle P^2 \rangle
                         - \langle P \rangle^2\right) \label{chidef} \\
  g_r            &=& \frac{\langle P^4 \rangle}{\langle P^2 \rangle^2} - 3.
\EA

Firstly we calculated the ground state energy for strong couplings 
$g_{\rm eff}^2$ and compared with the exact calculation \cite{NPFV}.
The obtained results from the lattice agree with the analytical result.
For higher values of $\beta_g$ the lattice calculation agrees within 
$10-20~\%$ with the previous effective double site calculation for 
the energy per site\cite{NPFV}. The ultraviolet regularization
with $\Dtau$ influences the final lattice result. The energies are 
measured with high accuracy and practically do not depend on the
infrared cutoff, i.e. on the total lattice size for $N \geq 8$.
The situation for other variables is different.

In order to see the emergence of a massless phase, we investigated
the (connected) time correlation functions of the (trace of the) 
Polyakov line operators:
\BA
  K(0,\tau) = \frac1{N^4} \left\langle
              \sum_{\vec{b},\vec{b'}}
              \cos(\varphi(\vec b,0))
              \cos(\varphi(\vec b',\tau))
              \right\rangle - \langle |P| \rangle^2 ~.
\EA
The correlation masses are obtained from a fit of the time correlation
functions $K(0,\tau)\vert_{\tau=n\Dtau}$ to a parameterization taking 
the periodicity in time into account 
\BE
  \label{mcorr}
  K(0,n\Dtau) = c\,
    \mbox{cosh}\!\left[m\Dtau\left(n-\frac{N_T}2\right)\right] ~.
\EE

\begin{figure}[ht]
\centerline{
  \scalebox{0.47}{\includegraphics{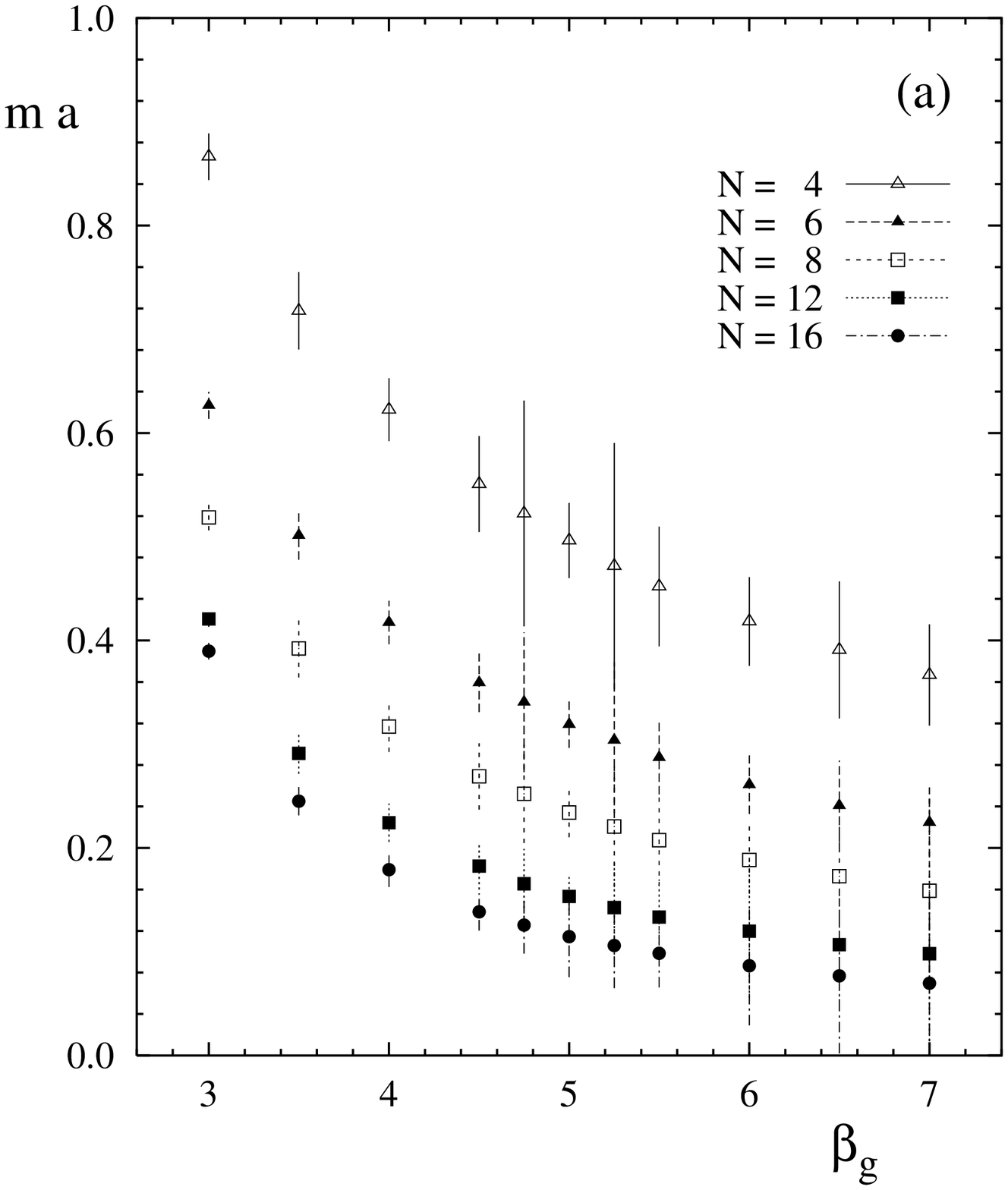}}~~
  \scalebox{0.47}{\includegraphics{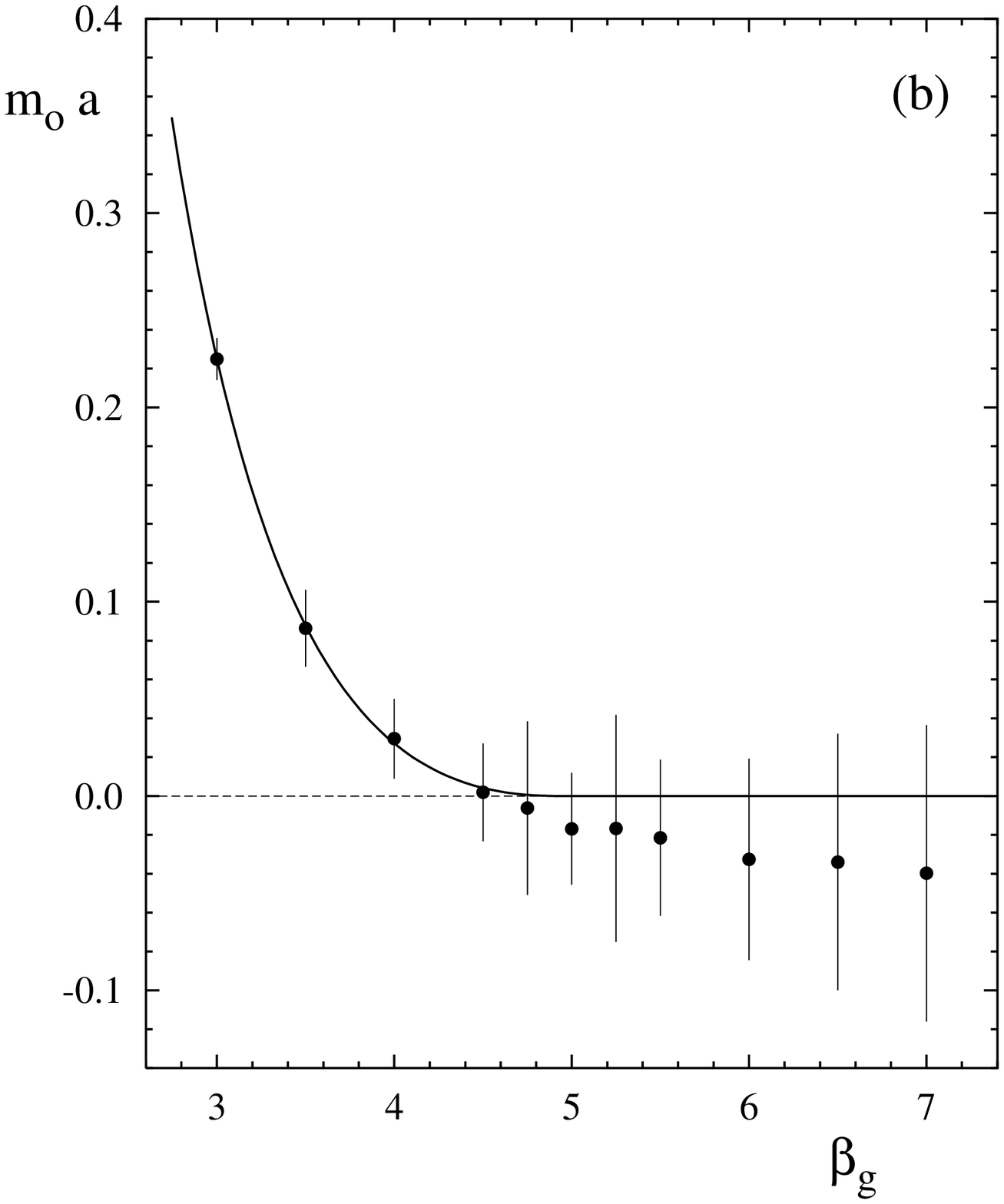}}
}
\caption{
  \label{FigMasses}
  Correlation masses $m$ for different effective couplings 
  $\beta_g = 1/g_{\rm eff}^2$: (a) -- for different lattice
  sizes $N$ from a fit of the measured correlation functions 
  $K(0,\tau)$ with the parameterization \Ref{mcorr}; (b) --
  in the limit $N \rightarrow \infty$ (see text). The solid line
  on (b) is the fit curve given by expression \Ref{mbeta}.
}
\end{figure}

The obtained masses for different lattice sizes are shown in
Fig.\ref{FigMasses}a. For each fixed $\beta_g$ we extrapolated
the mass to the limit $N\rightarrow\infty$ using a linear 
$1/N$ parameterization
\BE
  \label{mNfit}
  m(\beta_g,N) = m_0(\beta_g) + \frac{C(\beta_g)}{N} ~.
\EE
Our data for $m(\beta_g,N)$ can be fitted with this linear form 
\Ref{mNfit} for $\beta_g \gtsim 3$. The resulting dependence
of $m_0$ on $\beta_g$ is presented in Fig.\ref{FigMasses}b. The 
mass vanishes in the infinite volume limit near $\beta\approx5$,
which is a first guess for the critical coupling $\beta^*_g$.

In order to extract more exact information on the critical behavior
\BE
  \label{mbeta}
  m(\beta_g) \propto \left(\frac1{\beta_g}-\frac1{\beta_g^*}\right)^\nu ~.
\EE
from our data on small systems we have done a Finite Size Scaling 
(FSS) \cite{Binder,Barber} analysis of the variables $\absP$, $\chi$
and $g_r$, searching for the critical coupling $\beta_g^*$ in the 
interval $3 \leq \beta_g \leq 9$. We could determine the critical
indices $\beta$, $\gamma$ and $\nu$ using an approach employed for
$SU(2)$ gauge theory at the finite temperature transition by Engels
et al. \cite{Engels}. The general form of the scaling relations for
a variable $V$ ($V$ = $\absP$, $\chi$, $g_r$) is
\BE
  \label{FSSgen}
  V(t,N) = N^{\rho/\nu} F_V\left(t N^{1/\nu}, g_i N^{y_i}\right),
\EE
where $\rho$ is the corresponding critical index ($\beta$, $\gamma$, 0)
for the respective quantities and $t$ is the reduced inverse coupling 
$\beta_g$:
\BE
  \label{FSSt}
  t = \frac{\beta_g^* - \beta_g}{\beta_g}.
\EE
In practice eq. \Ref{FSSgen} is computed near $t=0$. Expanding up to first
order in $t$ and taking into account only the largest irrelevant exponent
$y_1\equiv-\omega$ one obtains
\BE
  \label{FSSone}
  V(t,N) = N^{\rho/\nu} 
  \left[c_0+\left(c_1+c_2 N^{-\omega}\right)tN^{1/\nu}+c_3 N^{-\omega}\right].
\EE

\begin{figure}[ht]
\centerline{
  \scalebox{0.47}{\includegraphics{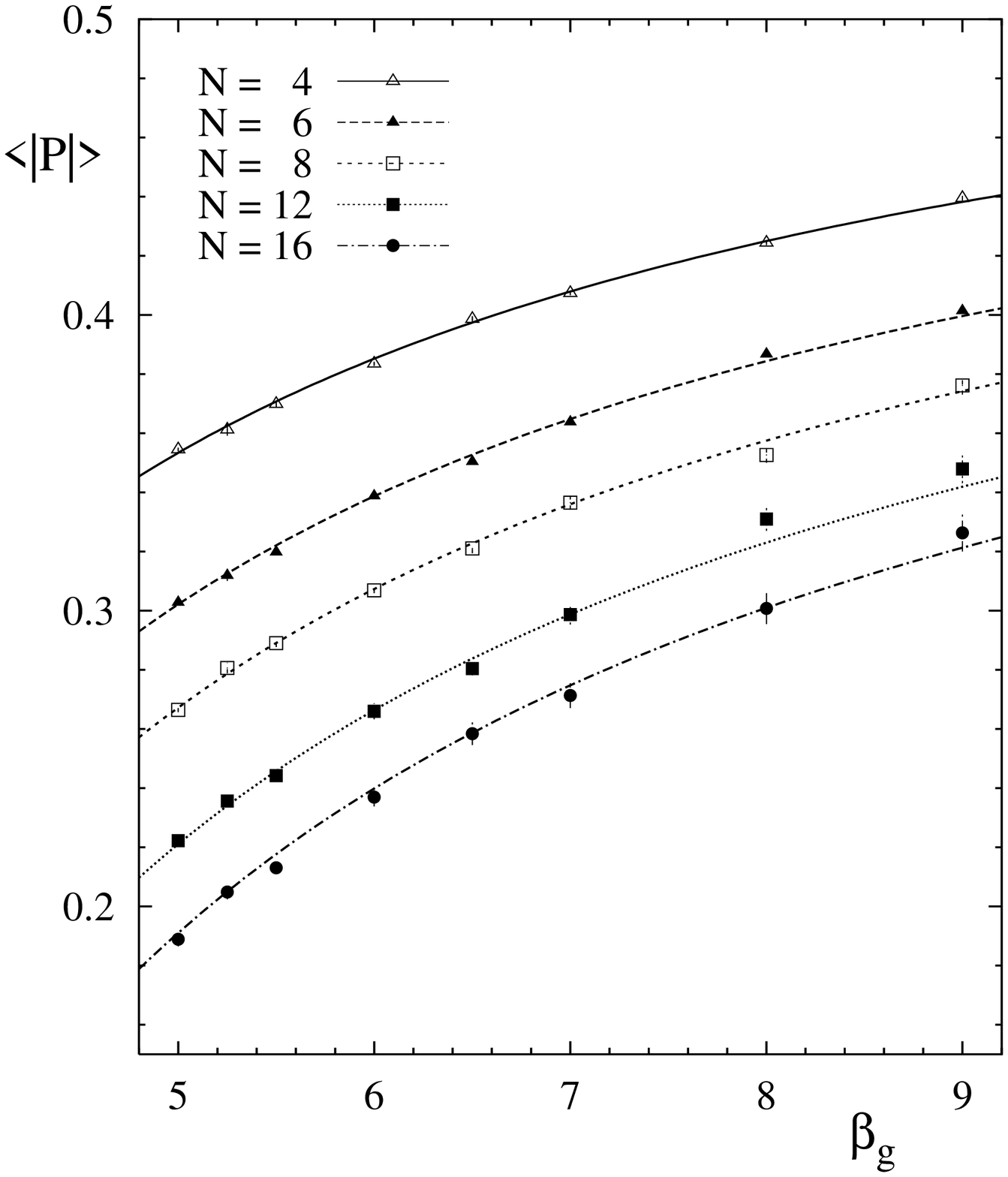}}~
  \scalebox{0.47}{\includegraphics{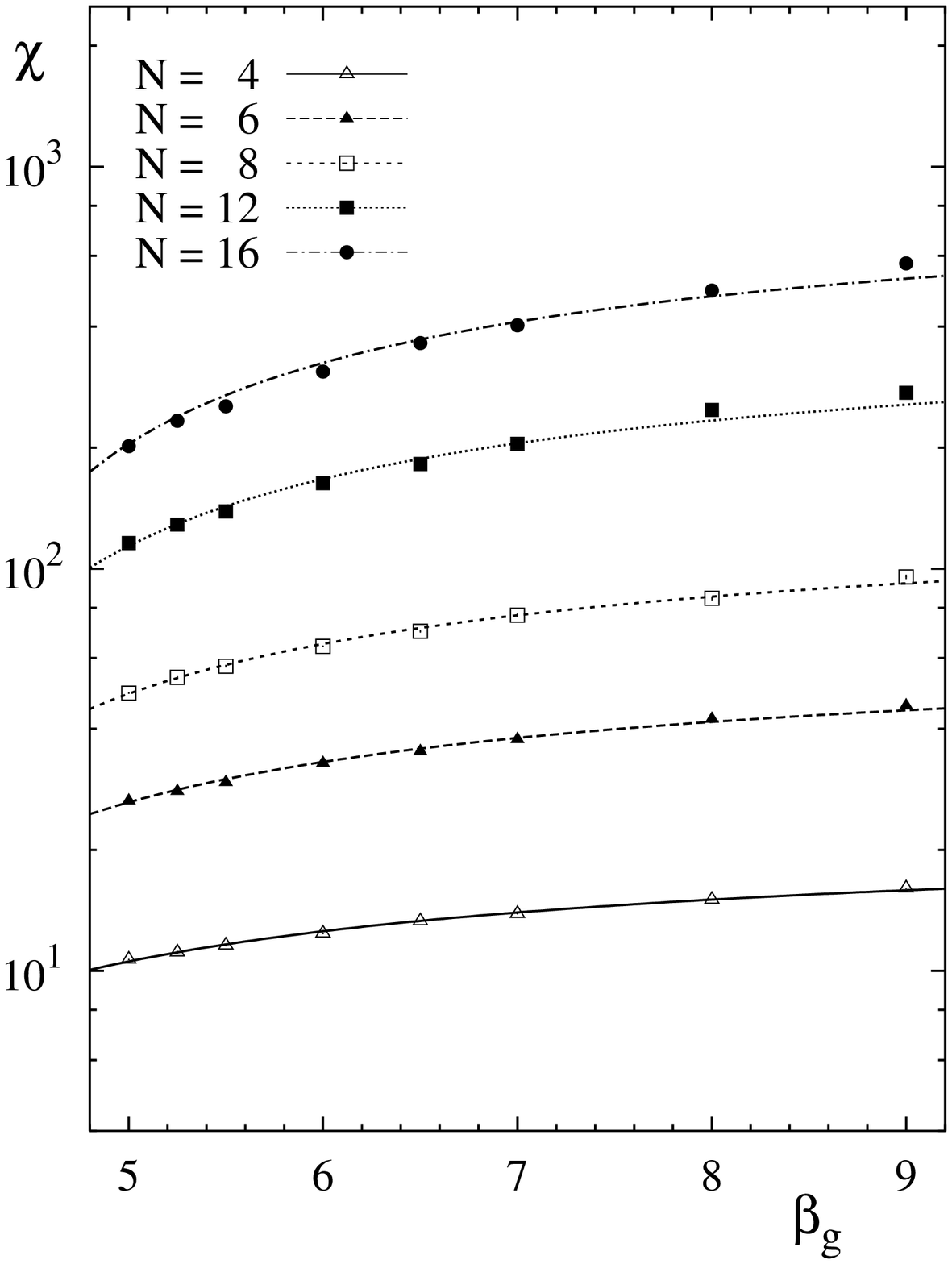}}
}
\caption{
  \label{FigFSS}
  Comparison of the theoretical parameterizations given in eqs.
  (\ref{FSSt}-\ref{FSSone}) for $\absP$ and $\chi$ with the
  lattice data results as a function of $\beta_g$ for different
  lattice sizes $N$.
}
\end{figure}

\begin{table}[ht]
\begin{center}
\begin{tabular}{|c|c|c|c|}\hline
         &             &                        &                \\[-2mm]
   $V$   &             & Transverse QCD $(2+1)$ &  Ising $(D=3)$ \\[-1mm]
         &             &                        &                \\\hline
         &             &                        &                \\[-3mm]
 $<|P|>$ & $\beta$     &    0.21 $\pm$ 0.06     &  0.3267(10)    \\
         & $\nu$       &    0.57 $\pm$ 0.06     &  0.6289(8)     \\
         & $\beta_g^*$ &    6.3  $\pm$ 1.1      &                \\[-3mm]
         &             &                        &                \\\hline
         &             &                        &                \\[-2mm]
 $\chi$  & $\gamma$    &    1.13 $\pm$ 0.36     &  1.239(7)      \\
         & $\nu$       &    0.52 $\pm$ 0.10     &  0.6289(8)     \\
         & $\beta_g^*$ &    5.7  $\pm$ 1.2      &                \\[-3mm]
         &             &                        &                \\\hline
\end{tabular}
\end{center}
\caption{
  \label{TabRes}
  Critical parameters for the transverse QCD near the light cone.
  The Ising indices are from ref.\cite{Ising}.
}
\end{table}

With this parameterization we analyze our data for lattice sizes 
$N=4,6,8,12,16$. The $\beta_g$ dependence in eq. \Ref{FSSone} is
valid near the critical point. The quality of the fit to the data
in different intervals on $\beta_g$ can serve as a guide to localize
the critical point $\beta_g^*$ \cite{Engels}. We find that in our case
this parameterization can be applied for $\absP$ and $\chi$ only for 
$\beta_g~\gtsim~5$: for smaller $\beta_g$ the $\chi^2/D.F.$ becomes 
too large. Finally we use the parameterization of eq. \Ref{FSSone}
in the interval $5 \leq \beta_g \leq 9$ (see  Fig.\ref{FigFSS}) where 
$\vert t\vert~\ltsim~0.3$ and an expansion of first order in $t$ 
still is applicable. In this $\beta_g$ interval the fits for $\absP$ 
and $\chi$ yield a $\chi^2/D.F\sim2$. To have a reliable error 
estimate for the extracted parameters we use the jackknife method. 
The results for $\beta_c^*$ and critical indices $\beta$, $\gamma$ 
and $\nu$ are presented in Table~1. We use $\omega=1$, but in fact
the results do not depend on $\omega$ in the interval 
$0.8\leq\omega\leq1.2$.

It should be noticed that the system frequently flips between the two
ordered states on a finite lattice. Therefore the expectation value 
$\langle P \rangle$ in equation \Ref{chidef} vanishes with good accuracy. 
Analogously to the treatment of the magnetization in the 3-dimensional 
Ising model, we should - instead of the expression for $\chi$ in eq.
\Ref{chidef} - define the susceptibility in the broken phase as
$\chi_{\rm broken}=N^3 \left(\langle P^2 \rangle - \absP^2\right)$. 
For the Ising case this susceptibility is supposed to converge towards
the correct infinite volume limit. Due to lack of statistics, however,
we did not use separate expressions for $\chi$ in the two phases.
Our data for $g_r$ are even much less accurate which excludes the
possibility to use them in the FSS analysis.

The values found for the critical indices $\beta$, $\gamma$ and $\nu$
are not far from those of the Ising model ($D=3$) (see Table~1). They 
are much less accurate but also in agreement with a high statistics 
analysis \cite{Engels} of the $SU(2)$ deconfinement transition 
at finite temperature. We interprete the behavior of the light
cone zero mode theory as a consequence of the underlying $Z(2)$
symmetry of the Hamiltonian. $Z(2)$ transformations correspond to 
reflections of the angle variables $\varphi$ around $\frac{\pi}{2}$.
The Hamiltonian and the measure of eq. (17) are invariant under these
transformations. The resulting critical behavior is common to
Ising like models. The critical coupling itself is a nonuniversal
quantity and we have found a rough value. More extended work with
the lattice Hamiltonian near the light cone is needed to clarify
the continuous transition further.

\section {Conclusions}

The scaling analysis gives indications that there is a second order
transition between a phase with massive excitations at strong coupling
and a phase with massless excitations in weak coupling. To reach a 
higher accuracy large scale simulations of the Hamiltonian zero mode
system are needed. In this context it is advisable to treat the coupled
system of three and two dimensional modes together. A calculation in the
epsilon expansion \cite{LZ} gives the zero of the $\beta$-function
as an infrared stable fixed point. This shows that the limit of large
longitudinal dimensions $L$ is well defined. Using the running coupling
constant $g^2_{\rm eff}$ of the zero mode system we have (\cf \Ref{mbeta})
\BE
  m\,a = \frac{1}{\zeta_0 g^{*\,2}_{\rm eff}}
    \left(g^2_{\rm eff} - g^{*\,2}_{\rm eff}\right)^\nu
\EE
with $g^{*\,2}_{\rm eff} = 0.17\pm0.03$ and $\nu = 0.56\pm0.05$ from 
our lattice calculations (\cf Table~\ref{TabRes}). The mass can be 
interpreted as an inverse correlation length  $ma = a/\xi$. When the
correlation length $\xi$ approaches the temporal size of the lattice 
$\xi \approx N_T\, a/2$ in the infrared limit we can identify $ma$ with
$a/L$, where $L$ is our infrared length scale. Therefore the effective
coupling $g^2_{\rm eff} = g^2 \eta \frac{L}{4a}$ (\cf \Ref{g2eff}) runs 
with $a/L$ in the following way
\BE
  \label{gevol}
  g^2_{\rm eff} = 
  g^{*\,2}_{\rm eff} 
  + \left(\zeta_0 g^{*\,2}_{\rm eff}\right)^{1/\nu}
    \left(\frac{a}{L}\right)^{1/\nu} ~.
\EE
The coupling to the three dimensional modes produces the usual
evolution of $g^2$ in $SU(2)$ QCD, where the coupling $g^2(a)$
at the lattice scale $a$ is related to the coupling $g^2_0$ 
defined at the infrared scale $\approx L$ as follows
\BE
  \label{gevola}
  g^2(a)= \frac{g^2_0}{1
        + \frac{g^2_0}{4\pi^2}\,\frac{22}{3}\,\log\frac{L}{a}}~.
\EE
Combining the eqs. \Ref{gevol} and \Ref{gevola} we can synchronize 
the approach to the light cone, {\it i.e.} the limit $\eta\rightarrow0$ 
with the continuum limit $a\rightarrow0$. The condition that the
three-dimensional evolution of $g^2$ has to be compatible with the
two-dimensional evolution of $g^2_{\rm eff}$ towards $g^{*\,2}_{\rm eff}$
yields that for $a\to0$ the light cone parameter $\eta$ approaches zero as
\BE
  \eta(a) \sim 
    \frac{g^{*\,2}_{\rm eff}}{\pi^2} \,
    \frac{22}{3} \, \frac{a}{L} \, \log\frac{L}{a} ~. 
\EE

This relation is similar to the naive scaling result deduced in section
$1$ from physics arguments besides logarithmic modification. Although at
the outset we had as parameters $N_L=L/a$ and $\eta$ in the Hamiltonian
we reduce this multiparameter problem to a problem with one single coupling
constant which can be chosen as $g$.

We can use the energy and the critical behavior of the Euclidean $2+1$
dimensional system to obtain its contribution to the ground state energy
of the system in $3+1$ dimensions. In our calculations the energy of the
zero mode Hamiltonian $\langle h \rangle$ near the critical coupling is
proportional to $N^2_\bot/\eta$. Using the scaling of $\eta$ determined
from the compatibility of 2-dimensional and 3-dimensional dynamics, we 
can eliminate $\eta$ in favor of $N_L = L/a$ and get a zero mode energy 
which grows linearly with the 3-dimensional volume. That means the zero 
mode dynamics becomes relevant for the full problem, {\it i.e.}
\BE
  \langle h a \rangle \sim N_L N^2_\bot ~.
  \label{meanha}
\EE
In reality the zero mode system is coupled to the $3+1$ dimensional system
the extension of which is large. So we expect that the couplings of the
$2+1$ dimensional system can get modified. Because of the universal dynamics
near the infrared fixed point this will not change the qualitative form of 
the volume dependence obtained in eq. \Ref{meanha}. The behavior of the
zero mode Hamiltonian has to be taken into account when we try to solve
for the infinite momentum frame solutions of the complete Hamiltonian. 
The discussion \cite {VFP, HMV} until now often centers on the order of
limits. In order to guarantee a meaningful limit of the near light cone
theory, one first has to take $L \rightarrow \infty$ and then one can 
take $\eta \rightarrow 0$. This is a workable procedure in solvable $1+1$
dimensional models. In QCD in $3+1$ dimensions it hardly can serve as a 
prescription. Numerical solutions have to be obtained with a finite cutoff
and the continuum limit can only be reached approximately via scaling 
relations. It is in this case that the role of zero mass excitations will
crucially enter the physics. As one sees above the synchronization of the
continuum limit with the light cone limit leads to results which are 
independent of the near light cone parameter $\eta$. The independence of
the ground state energy near the critical coupling gives us back Lorentz
invariance which states that the physics should not depend on the reference
frame. The vacuum energy is not allowed to depend on $\eta$. There are 
dynamical zero modes in near light cone QCD which contain the physics of
the nonperturbative QCD vacuum, the physics of the gluon and quark 
condensates. The nontrivial structure of the nonabelian gauge theory
enters in our calculation through the kinetic operator of the rotators
which at finite resolution have a finite mass gap vanishing only in the
continuum limit. 

The near light cone description of QCD based on a modified axial gauge 
$\partial_-A_-=0$ yields a very natural formulation of high energy scattering.
Polyakov variables near the light cone have already entered the calculations
of high energy cross sections in two different ways: For the geometrical size
of cross sections in ref.\cite{DGKP,N} the correlations between Polyakov 
lines along the projectile and target directions are important. For this
situation an interpolating gauge may be useful. For the energy dependence
of cross sections an evolution equation of Polyakov line correlators 
approaching the light cone has been discussed in ref. \cite{Bal} which
generalizes the BFKL- equation. This situation is close to the treatment
in this paper. Now the decisive step is to back up our theoretical work
on effective transverse QCD dynamics near the light cone with 
phenomenological consequences for high energy scattering. 

\section*{Appendix}

A choice of $\Dtau$ is proposed by comparing Monte-Carlo weights
$W_t$ and $W_s$ which link nearest neighbors in time and space
respectively
\BA
  W_t(\varphi_1,\varphi_2) &=& 
    \langle \varphi_1(x,y,\tau+1) \vert e^{-\Dtau h^{\rm e}} \vert
            \varphi_2(x,y,\tau  ) \rangle \,
    \sqrt{J(\varphi_1) J(\varphi_2)} ~, \\
  W_s(\varphi_1,\varphi_2) &=&
    \langle \varphi_1(x+1,y,\tau) \vert e^{-\Dtau h^{\rm m}} \vert
            \varphi_2(x  ,y,\tau) \rangle ~.
\EA
\begin{figure}[ht]
\vskip -15mm
\centerline{
  \scalebox{0.3}{\includegraphics{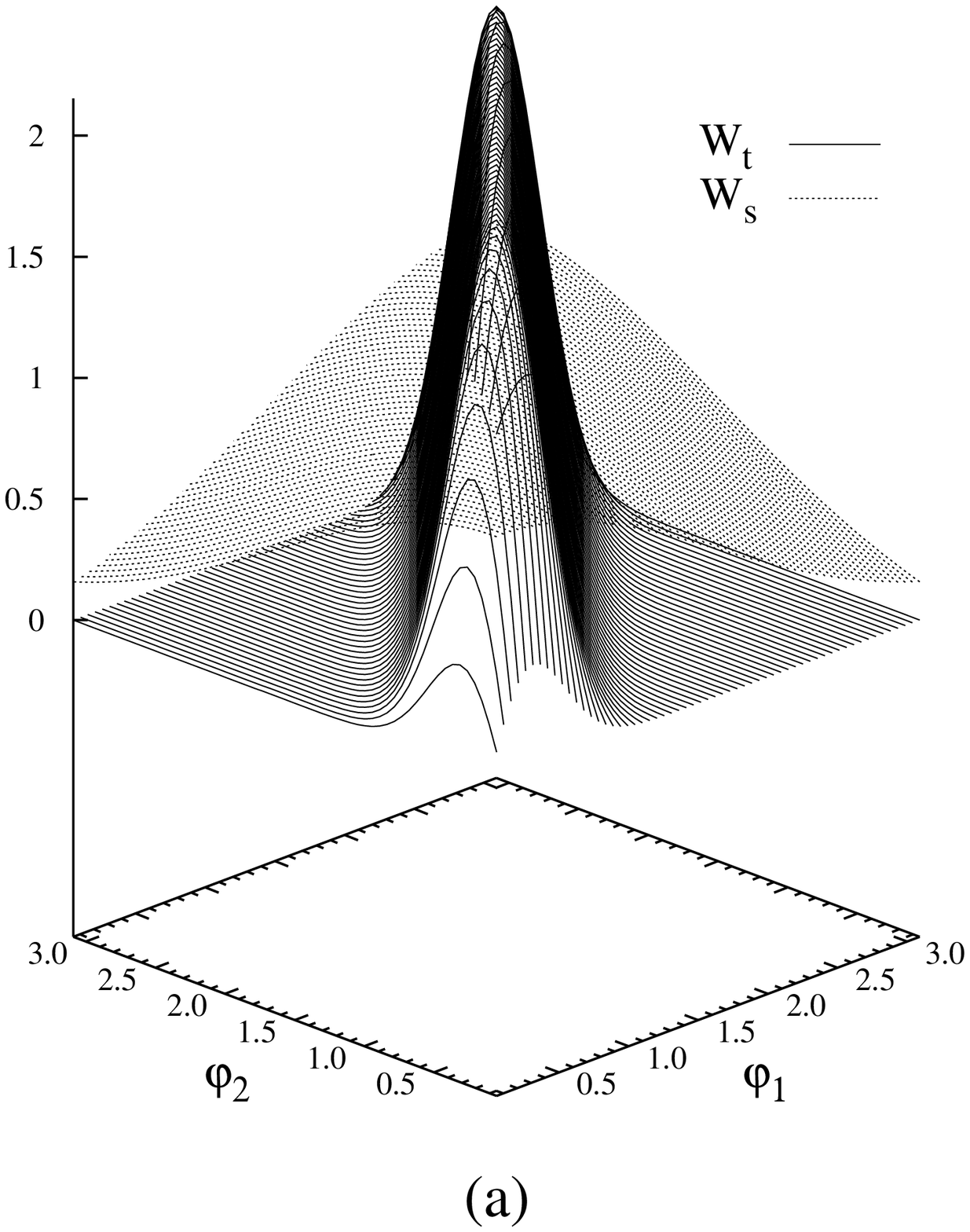}}~~
  \scalebox{0.3}{\includegraphics{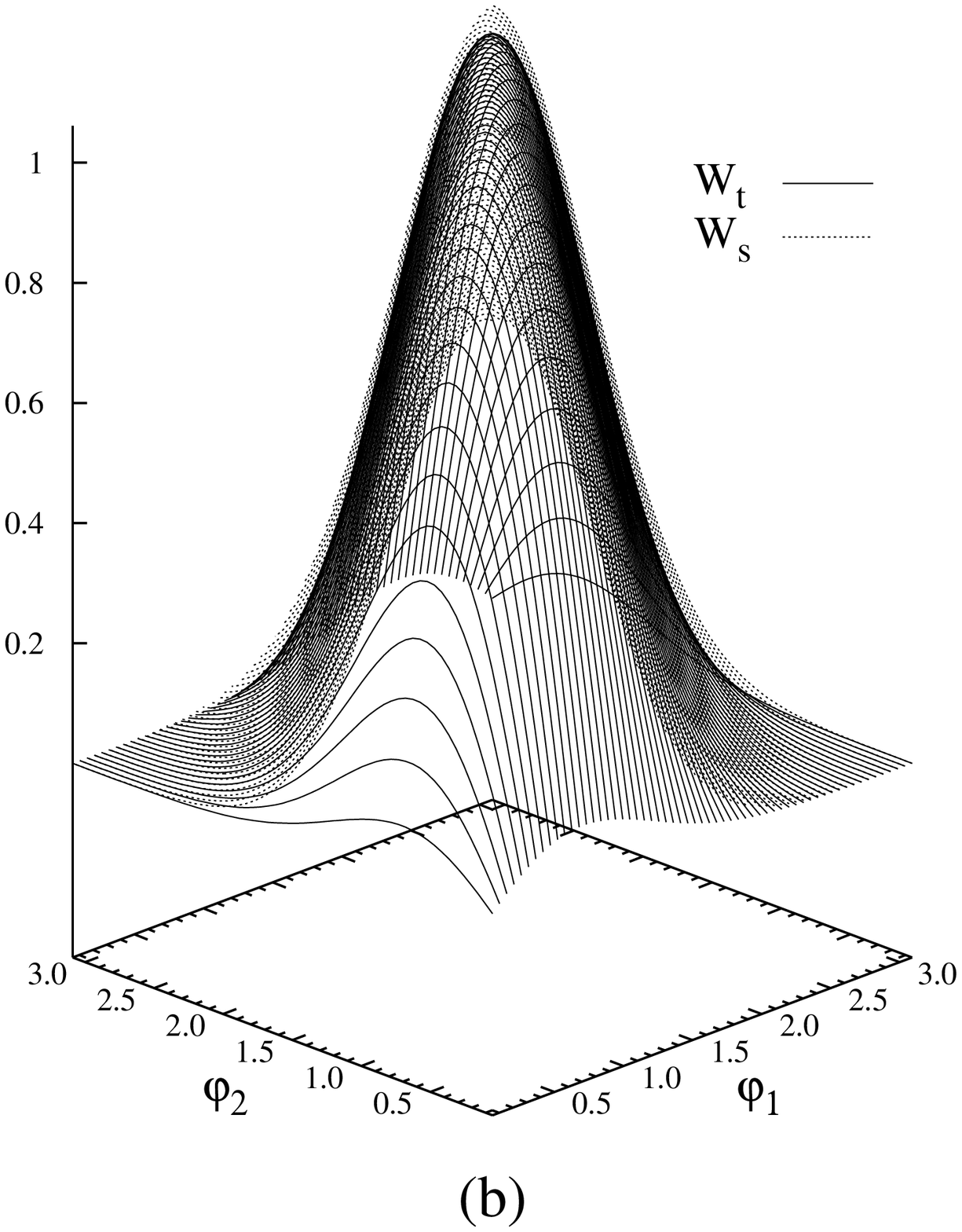}}~~
  \scalebox{0.3}{\includegraphics{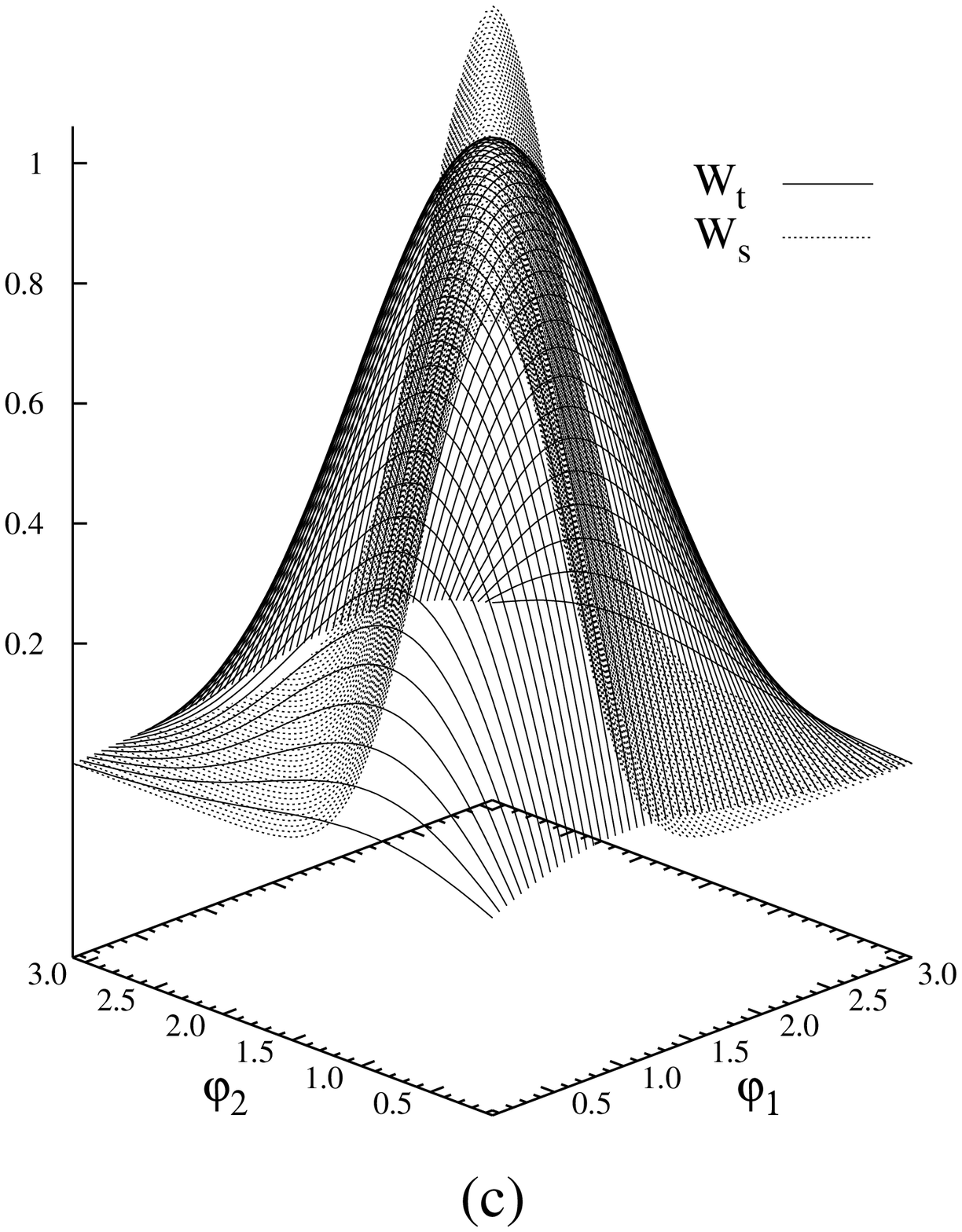}}
}
\vskip -7mm
\caption{
  \label{FigWlinks}
  Weights $W_t(\varphi_1,\varphi_2)$ of time-like and 
          $W_t(\varphi_1,\varphi_2)$ of space-like links
           for different $\Dtau$.
  Figures (a,b,c) correspond to $\Dtau = a/10, a/2, a$.
}
\end{figure}

The probability distribution for the update of the field 
$\varphi(x,y,\tau)$ is given by the product of weights for all
links connecting it to neighbors in time and space. So the 
updating procedure for $\varphi$ is more effective when $W_t$
and $W_s$ have similar distributions near the angle $\varphi=%
\pi/2$ where the Jacobian has its maximum. For $\Dtau=a/10$ 
(\cf Fig.~\ref{FigWlinks}a) the spatial weights $W_s$ have 
a much wider distribution than the timelike weights $W_t$, 
whereas for $\Dtau=a$ (\cf Fig.~\ref{FigWlinks}c) the 
situation is reverse. For the intermediate case $\Dtau=a/2$ 
(\cf Fig.~\ref{FigWlinks}b) the widths of the distributions
are very similar, therefore the updating procedure is optimal.


\end{document}